\documentclass{PoS}

\usepackage{amsfonts,amsmath,amssymb,graphicx,subfigure,braket}

\newcommand{\p}{\partial} 

\newcommand{\vp}[0]{\varphi} 

\renewcommand{\d}{\ensuremath{\mathrm{d}}}

\title{From unphysical gluon and ghost propagators to physical glueball propagators (in the Gribov-Zwanziger picture): a not so trivial task?}

\ShortTitle{From unphysical gluon and ghost propagators to physical glueball propagators}

\author{\speaker{David Dudal} and Nele Vandersickel\\
        Ghent University, Department of Physics and Astronomy, Krijgslaan 281-S9, 9000 Gent, Belgium\\
        E-mail: \email{david.dudal@ugent.be,nele.vandersickel@ugent.be}}

\author{Laurent Baulieu\\
Theory Division, CERN, 1211-Geneve 23, Switzerland \& LPTHE, Universit\'{e}s Pierre et Marie Curie, 4 place Jussieu, F-75252 Paris Cedex 05, France\\
        E-mail: \email{baulieu@lpthe.jussieu.fr}}

\author{Silvio P.~Sorella and Marcelo S.~Guimar\~{a}es\\
Departamento de F\'{\i }sica Te\'{o}rica, Instituto de F\'{\i }sica, UERJ - Universidade do Estado do Rio de Janeiro, Rua S\~{a}o Francisco Xavier 524, 20550-013 Maracan\~{a}, Rio de Janeiro, Brasil\\        E-mail: \email{sorella@uerj.br,msguimaraes@uerj.br}
        }

\author{Markus Q.~Huber\\
Theoretisch-Physikalisches Institut, Friedrich-Schiller-Universit\"at Jena, Max-Wien-Platz 1, 07743 Jena, Germany\\
        E-mail: \email{markus.huber@uni-jena.de}}

\author{Orlando Oliveira\\
Departamento de F\'{\i}sica, Universidade de Coimbra, P-3004-516 Coimbra, Portugal\\
        E-mail: \email{orlando@teor.fis.uc.pt}}

\author{Daniel Zwanziger\\
New York University, New York, NY 10003, USA\\
        E-mail: \email{daniel.zwanziger@nyu.edu}}

\abstract{During recent years, a good agreement was found between the analytical
derivation and the numerical simulation of the Landau gauge gluon and
ghost propagators. We mention the Schwinger-Dyson and Gribov-Zwanziger
formalism for the analytical work. Although the agreement between several approaches is nice, these
propagators do not correspond to the relevant physical degrees of
freedom. In the case of pure gauge theories, one should start to study
the glueball correlators. We shall try to explain why it looks like a
hard challenge to go from the unphysical to the physical propagators in
the case of the Gribov-Zwanziger theory (but similar conclusions might
hold for other approaches giving similar propagators).
}

\FullConference{Light Cone 2010: Relativistic Hadronic and Particle Physics\\
		 June 14-18, 2010\\
		 Valencia, Spain}

\begin{document}

\section{Introduction}
QCD is a strongly coupled theory, where nonperturbative physics plays a crucial role. As such, it is hard to handle analytically. A way out to investigate nonperturbative physics is model building (e.~g.~by using holographic QCD models, effective models like the PNJL model or others). Another extremely powerful tool is simulating QCD on a finite lattice. One can also attempt to quantize the theory in the continuum and try to get as good as possible information out of this by a variety of techniques. The latter philosophy shall also be employed in this work.

With lattice QCD, expectation values of gauge invariant operators can be computed without the need of gauge fixing. The eventual numerical estimates for physical quantities like a particle's mass are in good agreement with experimental data. The merit of lattice QCD is that it can also provide us with physical information in theories which do not appear in nature. Some famous examples are QCD with a number of colors other than 3 or QCD without quarks (gluodynamics). In the latter case, the physical spectrum supposedly exists of colorless pure glue states, the glueballs, see \cite{Mathieu:2008me,Chen:2005mg} and references therein.

We recall the classical $SU(N)$ Yang-Mills action in $d=4$ Euclidean space time,
\begin{eqnarray}
S_{YM}=\frac{1}{4}\int \d^4x F_{\mu\nu}^a F_{\mu\nu}^a\,.
\end{eqnarray}
This action possesses an enormous local invariance w.r.t.
\begin{eqnarray}
  A_\mu &\rightarrow& A_\mu^S~=~S^+\p_\mu S+S^+A_\mu S\qquad S\in SU(N)\,,
\end{eqnarray}
or in infinitesimal form
\begin{eqnarray}
A_\mu^a&\to~& A_\mu^a + D_{\mu }^{ab}\omega^b\,,\qquad D_{\mu
}^{ab}\equiv
\partial _{\mu }\delta ^{ab}-gf^{abc}A_{\mu }^{c}\,.
\end{eqnarray}
We need to reduce this enormous overcounting of physically equivalent gauge configurations by fixing a gauge. In principle we have a complete freedom to do so. Usually, one can pick a gauge suitable for the problem under study. However, there is an important restriction, as one should be assured that the eventual gauge fixed theory needs to be renormalizable. Not every gauge belongs to the class of renormalizable gauges.

We shall now focus on one particular example of a renormalizable gauge, viz.~the Landau gauge, $\p_\mu A_\mu=0$. This is a very popular gauge in the continuum, as it has many nice (quantum) properties \cite{Piguet:1995er}. According to the Faddeev-Popov procedure, the gauge fixed action reads
\begin{eqnarray}
S_{YM}+S_{gf}&=&\int \d^{4}x\;\left(\frac{1}{4}F_{\mu\nu}^2+ b^{a}\partial_\mu A_\mu^{a}
+\overline{c}^{a}\partial _{\mu } D_{\mu }^{ab}c^b\right)\,.
\end{eqnarray}
This gauge fixed action no longer enjoys a local gauge invariance. However it gets replaced by the equally powerful nilpotent BRST symmetry, $s (S_{YM}+S_{gf}) = 0$,
\begin{eqnarray}
sA_{\mu }^{a} =-D_{\mu }^{ab}c^{b}\,,  \qquad sc^{a} =\frac{g}{2}f^{abc}c^{b}c^{c} \,,\qquad s\overline{c}^{a} =b^{a}\,, \qquad sb^{a} =0\,,  \qquad s^2=0\,,
\end{eqnarray}
which can be used to prove the perturbative renormalizability and perturbative unitarity of the theory.

\section{Gauge (Gribov) copies and the Gribov-Zwanziger (GZ) approach}
During the Faddeev-Popov procedure, it is always tacitly assumed that there is one and only one solution to the gauge fixing condition. Gribov was the first to realize this to be wrong and constructed explicit examples in his seminal work \cite{Gribov:1977wm}. If we take $A_\mu$ in the Landau gauge, $\p_\mu A_\mu=0$, and consider an (infinitesimal) gauge transformation, $A_\mu'=A_\mu+D_\mu\omega$, then we quickly see that $\p_\mu A_\mu'=0$ if $\p_\mu D_\mu \omega=0$. Apparently, we encounter gauge copies if the Faddeev-Popov operator
\begin{equation}
M^{ab}=-\p_\mu D_\mu^{ab}
\end{equation}
has zero modes. In order to exclude these copies from the path integral, Gribov proposed to restrict the integration to  the Gribov region $\Omega$ where $\p A=0$ and $M>0$. This $\Omega$ corresponds to local minima of the functional $\int \d^4x A_\mu^2$ along the gauge orbits. This is already an improvement of the original Faddeev-Popov quantization procedure. The question turns out to be how to implement this kind of restriction to $\Omega$ in the continuum formulation? Gribov and later on Zwanziger \cite{Zwanziger:1989mf} worked out this problem and proved many properties of the region $\Omega$, for example that every gauge orbit passes through $\Omega$ \cite{Dell'Antonio:1991xt}. We should however warn that $\Omega$ still contains copies, not related to zero modes of the Faddeev-Popov operator \cite{vanBaal:1991zw}.

After a lengthy analysis, Zwanziger was able to implement the restriction to the Gribov region $\Omega$ to all orders by means of a local action, known as the Gribov-Zwanziger (GZ) action. The eventual GZ partition function becomes \cite{Zwanziger:1989mf}
\begin{equation}
Z_{FP} = \int [\d A][\d c][\d \overline c][\d b][\d \varphi][\d \overline \varphi] [\d \omega ][\d \overline \omega] e^{-S_{GZ}}\,,
\end{equation}
with
\begin{eqnarray}\label{gz}
S_{GZ} &=&S_{YM}+S_{gf} + S_{\varphi \overline{\varphi}
\omega \overline{\omega}}+ S_{\gamma}\,,\nonumber\\
S_{\varphi \overline{\varphi} \omega \overline{\omega}} &=& \int
\d^{4}x\Bigl( \overline{\varphi }_{\mu }^{ac}\partial
_{\nu}\left(\partial _{\nu }\varphi _{\mu }^{ac}+gf^{abm}A_{\nu
}^{b}\varphi _{\mu}^{mc}\right) -\overline{\omega }_{\mu
}^{ac}\partial _{\nu }\left( \partial_{\nu }\omega _{\mu
}^{ac}+gf^{abm}A_{\nu }^{b}\omega _{\mu }^{mc}\right)\,,\nonumber\\
&&-g\left( \partial _{\nu }\overline{\omega }_{\mu}^{ac}\right) f^{abm}\left( D_{\nu }c\right) ^{b}\varphi _{\mu}^{mc}\Bigr)  \nonumber  \\
S_{\gamma}&=& -\gamma ^{2}g\int\d^{4}x\left( f^{abc}A_{\mu
}^{a}\varphi _{\mu }^{bc}+f^{abc}A_{\mu}^{a}\overline{\varphi }_{\mu
}^{bc} + \frac{4}{g}\left(N^{2}-1\right) \gamma^{2} \right)\,,
\end{eqnarray}
where extra bosonic ($\varphi$, $\overline\varphi$) and fermionic ($\omega$, $\overline\omega$) fields were introduced. The parameter $\gamma$ carries the dimension of mass and \emph{must} be self-consistently fixed to the nonzero solution of the following gap equation
\begin{equation}\label{gap}
\frac{\p E_{vac}}{\p \gamma}=0\,,
\end{equation}
commonly known as the horizon condition \cite{Zwanziger:1989mf}, thereby giving $\gamma\sim\Lambda_{QCD}$, a typical example of dimensional transmutation. It can be easily checked that for $\gamma=0$ one recovers the Faddeev-Popov theory. It is important to mention that this action \eqref{gz} defines a renormalizable theory see \cite{Zwanziger:1989mf,Dudal:2010fq} and references therein.

What about the BRST symmetry? One can naturally extend the BRST symmetry to the new fields\footnote{For $\gamma=0$, we then obtain a trivial extension of the usual Faddeev-Popov gauge theory.}
\begin{eqnarray}\label{GZb3b}
s\varphi _{\mu}^{ac} ~=~\omega_{\mu}^{ac}\,,\quad s\omega_{\mu}^{ac}~=~0\,,\quad s\overline{\omega}_{\mu}^{ac} ~=~\overline{\varphi }_{\mu}^{ac}\,,\quad s \overline{\varphi }_{\mu}^{ac}~=0~\,.
\end{eqnarray}
The symmetry of the original action under the BRST transformation is softly broken
\begin{eqnarray} sS_{GZ}=g\gamma^2\int \d^4x\left(f^{abc}A_\mu^a\omega_\mu^{bc}-(D_\mu^am c^m)(\overline{\varphi}_\mu^{bc}+{\varphi}_\mu^{bc})\right)\,.
\end{eqnarray}
With softly we mean that it is proportional to the mass parameter $\gamma^2$, thus it can be controlled at the quantum level. Very recently, an equivalent formulation of the GZ theory was given in which case the breaking is even converted into a linear breaking \cite{Capri:2010hb}.

Apparently, treating gauge copies \`{a} la GZ leads to a loss of the BRST symmetry. As such, the situation of how to define physical states also becomes less clear (see later).

\section{Why studying propagators?}
During the past decade, a lot of effort went into the investigation of the elementary gluon and ghost propagators. One might wonder why so much study is devoted to these gauge variant quantities, as these do not correspond to physically measurable quantities? Propagators are the basic building blocks of quantum field theory: they do describe the propagation of the elementary, albeit perhaps unphysical, degrees of freedom and they are the ``simplest'' objects to compute. In any Feynman diagram-based approach to QCD, propagators appear. This reaches far beyond perturbation theory, one needs only to think about Schwinger-Dyson, Bethe-Salpeter, sum rules, $\ldots$ approaches to QCD, which are all nonperturbative in nature.  In addition, since gluons are confined at low energy, we might expect to see something nontrivial already at the level of the propagators.

These propagators were also intensively studied using lattice simulations, in particular in the Landau gauge. This gauge is very suitable to be simulated, as it corresponds to searching for the minima of the functional $\int \d^4x A_\mu^2$ along the gauge orbits. As such, one gets the ``exact'' propagators and can compare those with analytical results in order to test the latter. If analytical and lattice results are in good agreement, we may have a certain degree of confidence that the analytical approximation scheme gives reasonably good propagators, which can then be used in other computations which depend heavily on knowledge of nonperturbative propagators.

In Figure 1, taken from \cite{Dudal:2010tf}, we show the lattice SU(3) Landau gauge gluon propagator
\begin{figure}[b]
   \centering
      \includegraphics[width=6cm]{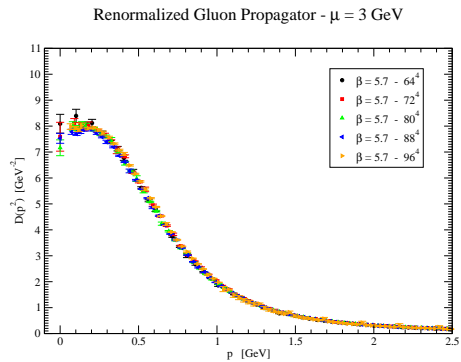}
     \caption{Gluon propagator renormalized at $\mu=3$ GeV.}
\end{figure}
and it is clear that there is no sign of a blow-up in the infrared, typically occurring when using perturbation theory in the massless Faddeev-Popov scheme. This is indicative of nonperturbative effects, and we shall try to motivate that these might be related to Gribov copies.

\section{The Refined Gribov-Zwanziger (RGZ) approach}
Using the action \eqref{gz}, it is readily verified that the tree level gluon propagator reads
\begin{equation}
    D(p^2)=\frac{p^2}{p^4+\lambda^4}\,,
\end{equation}
which is indeed infrared suppressed, although it vanishes, in contradistinction with the lattice Figure 1. We have set here $\lambda^4=2g^2N \gamma^4$. These effects persist upon including loop corrections \cite{Gracey:2006dr}. For the ghost propagator $G(p^2)$, one can prove at any order \cite{Zwanziger:1989mf} or compute explicitly \cite{Gracey:2005cx} that by invoking the gap equation \eqref{gap}
\begin{equation}
    p^2G(p^2)\sim\frac{1}{p^2}\,,\qquad\text{for}~ p^2\sim0\,,
\end{equation}
which again seems to be at odds with large volume lattice data \cite{Cucchieri:2008fc}.

Apparently, something is missing in the Gribov-Zwanziger formulation as standing. In order to overcome this, in \cite{Dudal:2007cw,Dudal:2008sp} extra dynamical effects due to nonperturbative $d=2$ condensates were taken into account. We recall that the $d=2$ condensate was popularized in the last decennium, thanks to works like \cite{Gubarev:2000nz,Verschelde:2001ia,Boucaud:2001st}. As a result of the analysis, one finds a gluon propagator of the form \cite{Dudal:2007cw,Dudal:2008sp}
\begin{eqnarray}\label{gl}
D(p^2)=\frac{p^2+M^2}{p^4+(m^2+M^2)p^2+M^2m^2+2g^2N\gamma^4}\,,
\end{eqnarray}
whereby $m^2$ and $M^2$ are mass scales corresponding to condensates, in particular
\begin{eqnarray}\label{masses}
    m^2\sim \braket{A^2}\,,\qquad M^2\sim \braket{\overline\varphi\varphi-\overline\omega\omega}\,.
\end{eqnarray}
We observe that $D(p^2)$ is still infrared suppressed, but $D(0)\neq0$ thanks to the presence of $M^2$. Hence, the gluon propagator lattice data is already qualitatively reproduced. For the ghost, one finds \cite{Dudal:2007cw,Dudal:2008sp} $G(p^2)\sim \frac{1}{p^2}$ for $p^2\sim0$, again consistent with lattice data.

Having found a reasonable qualitative agreement, one might wonder whether the lattice data for the gluon could also be quantitatively fitted with a propagator of the form \eqref{gl}. This was tested in \cite{Dudal:2010tf}. The fit with $m^2=0$ did not work out well, indicating that $\braket{A^2}$ is of importance. The following continuum extrapolated values were reported
\begin{equation}\label{scale1}
M^2 = 2.15 \pm 0.13~\text{GeV}^2\,,\qquad m^2 = - 1.81 \pm 0.14~\text{GeV}^2\,,\qquad 2 g^2 N \gamma^4 = 4.16 \pm 0.38~\text{GeV}^4\,,
\end{equation}
leading to the fits displayed in Figure 2 for the gluon propagator and its form factor.
\begin{figure}[t]
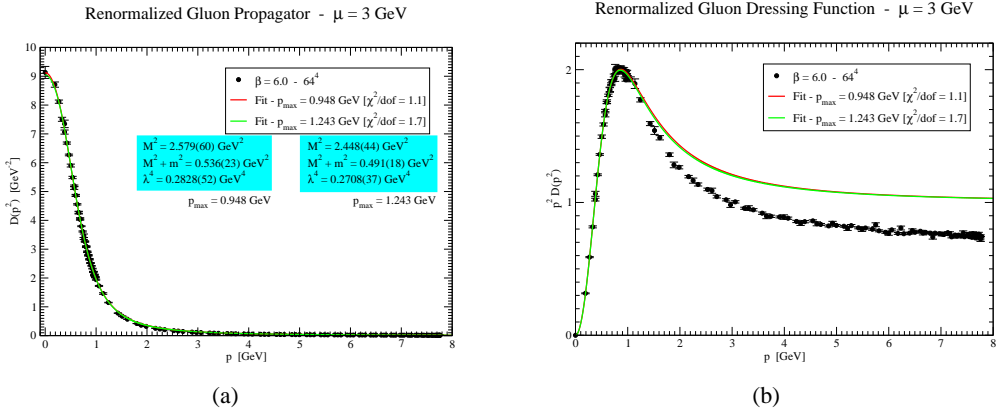

  \begin{center}
  \subfigure[]{\includegraphics[width=6cm]{procfig2.eps}}
    \hspace{1cm}
    \subfigure[]{\includegraphics[width=6cm]{procfig3.eps}}
     \end{center}
  \caption{Fit to the lattice gluon propagator (left) and form factor (right).}
\end{figure}
We see that the fit works out well up to $p\sim 1.5 \text{GeV}$. As a byproduct of this analysis, an estimate for the $d=2$ gluon condensate was obtained,
\begin{equation}
\braket{g^2A^2}_{\mu=10\text{GeV}}\approx 3 \text{GeV}^2\,,
\end{equation}
which is in the same ballpark as other, rather independent, approaches, see \cite{Dudal:2010tf} for references.

It is apparent that the RGZ framework seems to be able to describe quite well the nonperturbative (gluon) propagator. One is thus lead to believe that RGZ can be a good starting point to study nonperturbative aspects of Landau gauge QCD. Future work should be devoted to try to compute the scales in $D(p^2)$ in a clean analytical fashion using an effective potential approach.  Recently, in \cite{Gracey:2010cg} an even more general RGZ setting was proposed. We conclude this first part of our talk by drawing attention to similarly good-looking results for the two basic propagators in other approaches obtained by solving Schwinger-Dyson equations, see some of the other proceedings and \cite{Fischer:2008uz,Binosi:2009qm}.

\section{Glueballs in the (R)GZ approach and the concept of $i$-particles}
There is however more to life than gluon and ghost propagators. As the latter are believed to be unphysical, one should look what the physical degrees of freedom could be, such that these can be described by the (R)GZ theory. Therefore, we need to look at glueball correlation functions. The big question is how to proceed from the ``simple'' gluon/ghost to the ``less simple'' glueball?

It turns out to be useful to consider two sets of variables, being the old variables (gluons, ghosts and extra GZ fields), which are useful for renormalization issues, and a set of new variables, the so-called $i$-particles \cite{Baulieu:2009ha}, useful for spectral issues.

Let us first introduce these $i$-particles. We notice that there are 2 complex (conjugate) gluon ``masses'' in the propagator \eqref{gl}, given certain bounds on mass scales, which are for example fulfilled by the numbers \eqref{scale1}. A set of 2 \emph{cc} masses immediately leads to a tree level gluon positivity violation (which is also seen on the lattice), meaning that the gluon cannot be a physical excitation \cite{Osterwalder:1974tc}. One can believe this is reflective of gluon confinement \cite{Zwanziger:1991gz}. We also observe that there is a field mixing in the tree level RGZ action
\begin{eqnarray}
\int \d^4x\; \left( \frac{1}{2} A^a_{\mu} (-\partial^2\delta_{\mu\nu}-\p_\mu\p_\nu) A^a_{\nu} + {\bar \vp}^{ab}_{\mu}(-\partial^2)\vp^{ab}_{\mu} +\gamma^2\,g\,f^{abc}A_\mu^{a}(\vp_\mu^{bc}-{\bar\vp}_\mu^{bc}) +b^a \p_\mu A_\mu^a\right)\,.
\end{eqnarray}
These $cc$ masses are a bit hidden in these variables; we shall also set $m^2=0$ for the remainder of this talk. By using a set of linear transformations \cite{Baulieu:2009ha}, the foregoing tree level action can however be recast in the following form
\begin{eqnarray*}
\int \d^4x  \left[ \frac{1}{2} {\lambda}^{a}_{\mu} \left( -\partial^2  + \frac{M^2 - i \sqrt{ 4 \lambda^4 - M^2 } }{2} \right)  {\lambda}^{a}_{\mu} + \frac{1}{2} {\eta}^{a}_{\mu}\left( -\partial^2 + \frac{M^2 + i \sqrt{ 4 \lambda^4 - M^2}}{2}   \right)   {\eta}^{a}_{\mu}  +\text{rest}\right]\,.
\end{eqnarray*}
Clearly, the new fields $\lambda_\mu^{a}$ and $\eta_\mu^a$ have \emph{cc} masses. They contain the gluon fields, which are unavoidably mixed with the (R)GZ content due to $\gamma\neq0$. We call these novel fields the $i$-particles of the (R)GZ theory. The question how to describe glueballs in the (R)GZ context is still unanswered.

A first potential pitfall is how to define a physical subspace without BRST invariance?  These $i$-particles are obviously unphysical, which can be called ``confinement'' if one is of a very optimist nature. But gluon confinement is of course more. We should be able to define a physical subspace of purely gluonic states (i.e.~glueballs), which should not decay into unphysical gluons/ghosts/$\ldots$. It looks like we need a symmetry to define such a subspace, thereby expelling the unphysical stuff. This smells like a BRST symmetry application, but the GZ action breaks BRST as we have seen already. In \cite{Dudalextra}, it was shown how to construct a local albeit not nilpotent BRST symmetry of an equivalent version of the GZ theory, but it remains unclear if this symmetry is sufficiently strong to define a physically sensible set of glueball operators.

A second pitfall concerns the renormalization of a suitable glueball operator. As a glueball is a kind of bound state of gluons, we need a suitable (local) composite operator, whose quantum numbers correspond to the glueball state under investigation. As glueballs are physical, we expect gauge invariant composite operators\footnote{Or more precisely, quantum BRST cohomology classes, if we have a nilpotent BRST operator.}. Again, the loss of the quantum version of the BRST symmetry seems problematic. Nevertheless, in case of the scalar glueball, it was shown in \cite{Dudal:2009zh} that a renormalization group invariant extension of the classically gauge invariant operator $F_{\mu\nu}^2(x)$ can be constructed in (R)GZ. As expected, there is operator-mixing into\footnote{EOM-terms stand for terms proportional to the equations of motion.}
\begin{equation}
O= F_{\mu\nu}^2+s(d=4~\text{operators})+\gamma^2(d=2~\text{operators})+\text{EOM-terms}\,.
\end{equation}
Notice that, unlike in usual QCD,
\begin{equation}
    \Braket{O(x) O(y)}\neq \Braket{F_{\mu\nu}^2(x) F_{\rho\sigma}^2(y)}\,,
\end{equation}
the reason being the BRST breaking $\sim \gamma^2$. This means that the mixing terms do influence the correlator.

How should a general glueball operator $O(x)$ look like in the (R)GZ world? $O(x)$ should be renormalizable, and for $\gamma=0$, we expect to find back the original QCD cohomology output, thus something of the form
\begin{eqnarray}
O_{\gamma=0}(x)&=& \text{gauge invariant operator} + s(d=4~\text{operator})+\text{EOM-terms}\,.
\end{eqnarray}
Hence, for $\gamma\neq0$, we seem to be driven to
\begin{eqnarray}\label{vb}
O_{\gamma\neq0}(x)&=& \text{gauge invariant operator} + s(d=4~\text{operator})\nonumber\\ &&+ \gamma^2 (d=2~\text{operator})+\text{EOM-terms}\,.
\end{eqnarray}
The discussion on renormalization of classically gauge invariant operators is most easily given in the old variables. The story changes when we want to check whether the operator generates a physical two-point function. In order to speak about a physical propagator $\Delta(p^2)= \braket{O(p)O(-p)}$, $D(p^2)$ must have a decent spectral representation\footnote{We work in Euclidean space.},
\begin{equation}\label{pre1}
 \Delta(p^2)=\frac{Z}{p^2+m_*^2}+\int_{\tau_0}^\infty \frac{\rho(t)}{t+p^2}\d t\,,\qquad p^2\in\mathbb{C}
\end{equation}
so that $\Delta(p^2)$ displays a branch cut only along the negative axis, with positive discontinuity $\rho(t)$, given by
\begin{equation*}\label{pre2}
    \rho(t)=\frac{1}{2\pi i}\lim_{\epsilon\to 0^+}\left[\Delta(-t-i\epsilon)-\Delta(-t+i\epsilon)\right]\,.
\end{equation*}
This positivity can be easily understood from the optical theorem since $\rho$ is proportional to the cross section, which ought to be positive. The $m_*^2>0$ correspond to physical particle masses, while $\tau_0>0$ corresponds to the multiparticle-threshold.

As demonstrated in \cite{Baulieu:2009ha}, the $i$-particles are very suited to derive the spectral representation of e.g.~the correlation function built with $F_{\mu\nu}^2$. At lowest order, we can work in the quadratic tree level (Abelian) approximation, in which case $F_{\mu\nu}^2$, rewritten in $i$-particle field strengths, reads
\begin{equation}
o(x)=\frac{1}{2}f_{\mu\nu}^af_{\mu\nu}^a=\lambda_{\mu\nu}^a\eta_{\mu\nu}^a+\frac{1}{2}\lambda_{\mu\nu}^a\lambda_{\mu\nu}^a+\frac{1}{2}\eta_{\mu\nu}^a\eta_{\mu\nu}^a\,.
\end{equation}
A priori, it is unclear how to compute the spectral representation, as the Cutkosky cut rules are in principle only intended for use with real masses in Minkowski space. This problem can nevertheless be handled, but discussing this here would lead us to far. Let it suffice to mention that a few examples were worked out in \cite{Baulieu:2009ha}.

For $o_1(x)=\lambda_{\mu\nu}^a\eta_{\mu\nu}^a$, it turns out that
\begin{equation}
\Braket{o_1(p)o_1(-p)}=\int_{\tau_0}^{\infty}\frac{\rho(t)\d t}{t+p^2}\,,\qquad \text{with }\rho(t)\geq0\,.
\end{equation}
This is good news, as this represents a physical spectral representation.  On the contrary, $o_2(x)=1/2\lambda_{\mu\nu}^a\lambda_{\mu\nu}^a+1/2\eta_{\mu\nu}^a\eta_{\mu\nu}^a$ leads to
\begin{equation}
\Braket{o_2(p)o_2(-p)}=\int_{\text{curve in}~\mathbb{C}}\frac{\rho(t)\d t}{t+p^2}+cc
\end{equation}
thereby displaying cuts in the complex plane. Unfortunately, this also means that the tree level version of $F_{\mu\nu}^2$ itself, viz.~$f_{\mu\nu}^2=o_1+o_2$, leads to an unphysical correlator \cite{Baulieu:2009ha}.

It would then seem that only taking $\lambda_{\mu\nu}^a\eta_{\mu\nu}^a$ is a good choice, since
\begin{equation}
\lambda_{\mu\nu}^a\eta_{\mu\nu}^a=f_{\mu\nu}^2+\text{rest}\,,
\end{equation}
and it leads to a physical correlation function. However, this operator falls outside the class \eqref{vb}, and it therefore seems to be doubtful to be renormalizable, i.~e.~controllable beyond the tree level. In addition, we do not have any information yet on the spectral properties beyond tree level either.

We are currently investigating an operator of the kind
\begin{equation}\label{l}
    f_{\mu\nu}^2+s(\text{other operators})+\gamma^2(\text{other operators})\,,
\end{equation}
restricted to the Abelian level. If the spectral properties would turn out to be OK, the form \eqref{l} would allow to at least write down an extension of the operator to the quantum level. One can then try to investigate its renormalization and, if possible, its higher order spectral properties. This kind of operator \eqref{l} would also fit with the new BRST constructed in \cite{Dudalextra}. From the tree level results, if they are physically decent, one can also already extract information on glueball masses, perhaps along the lines of \cite{Capri:2010pg} where a GZ-like theory was studied in relationship with glueball-like operators.

We conclude that it is apparently very hard to accommodate good renormalization and good analyticity properties at the same time when it comes to the study of glueball operators. We believe this is not only of relevance to work in the (R)GZ context, but to all people active in propagator research: how can one go from the unphysical gluon/ghost propagators to a well-defined physical subspace of glueball operators, which can be controlled at the quantum level. In addition, one should also try to find reasonable estimates for the glueball masses \cite{Chen:2005mg}.
This looks like a very ambitious program, but we hope that it will stimulate a lot of new research in the coming years. It would also be interesting to find out whether the difference between GZ and RGZ plays a role when it comes to glueball properties. Also the issue of a BRST symmetry needs to be further clarified, even in the Schwinger-Dyson context due to the potentially subtle role played by boundary conditions \cite{Fischer:2008uz,Maas:2009se,Dudal:2009xh}.

\section*{Acknowledgments}
D.~Dudal wishes to thank the organizers for the kind invitation. D.~Dudal and N.~Vandersickel are supported by the Research-Foundation (Flanders). O.~Oliveira is supported by FCT under project CERN/FP/83644/2008. S.~P.~Sorella and M.~S.~Guimar\~{a}es gratefully acknowledge financial support of the Conselho
Nacional de Desenvolvimento Cient\'{\i}fico e Tecnol\'{o}gico (CNPq-Brazil),
the Faperj, Funda{\c{c}}{\~{a}}o de Amparo {\`{a}} Pesquisa do Estado do Rio
de Janeiro, the SR2-UERJ and the Coordena{\c{c}}{\~{a}}o de Aperfei{\c{c}}%
oamento de Pessoal de N{\'{\i}}vel Superior (CAPES) and the Latin American Center for Physics (CLAF).

\end{document}